\documentclass[aip,reprint,superscriptaddress,showpacs]{revtex4-1}
\usepackage{amsmath}
\usepackage{amssymb}
\usepackage[utf8]{inputenc}
\usepackage[english]{babel}
\usepackage{mathrsfs}
\usepackage{bm}

\begin{document}

\title{Finite-dimensional colored fluctuation-dissipation theorem for spin systems}

\author{Stam Nicolis}
\email{stam.nicolis@lmpt.univ-tours.fr}
\affiliation{CNRS-Laboratoire de Mathématiques et Physique Théorique (UMR 7350), Fédération de Recherche "Denis Poisson" (FR2964), Département de Physique, Université de Tours, Parc de Grandmont, F-37200, Tours, FRANCE}

\author{Pascal Thibaudeau}
\email{pascal.thibaudeau@cea.fr}
\affiliation{CEA DAM/Le Ripault, BP 16, F-37260, Monts, FRANCE}

\author{Julien Tranchida}
\email{julien.tranchida@cea.fr}
\affiliation{CEA DAM/Le Ripault, BP 16, F-37260, Monts, FRANCE}
\affiliation{CNRS-Laboratoire de Mathématiques et Physique Théorique (UMR 7350), Fédération de Recherche "Denis Poisson" (FR2964), Département de Physique, Université de Tours, Parc de Grandmont, F-37200, Tours, FRANCE}


\begin{abstract}
When nano-magnets are coupled to random external sources, their magnetization becomes a random variable, whose properties are defined by an induced probability density, that can be reconstructed from its moments, using the Langevin equation, for mapping the noise to the dynamical degrees of freedom.  When the spin dynamics is discretized in time, a general fluctuation-dissipation theorem, valid  for non-Markovian noise, can be established, even when zero modes are present. We discuss the subtleties that arise, when Gilbert damping is present and the mapping between noise and spin degrees of freedom is non--linear.

\end{abstract}


\maketitle

\section{Introduction}
For any system, in equilibrium with a bath, the fluctuation-dissipation relation (FDR) plays an important role in defining consistently its closure, since it relates the fluctuations of the subsystem of the dynamical degrees of freedom,  that one is, by definition, interested in, with the fluctuations of the degrees of freedom that are defined as uninteresting and are lumped under the term ``dissipation''. 

The essential reason behind this relation is that, for equilibrium situations, it is possible to define a probability measure on the space of states, with respect to which the average values, that enter in the FDR, can be unambiguously computed. So this can be modified, if the dynamical degrees of freedom are so affected by the immersion in the bath, that they must be replaced by others--the interaction with the bath leads to a phase transition and the equilibrium measure is not unitarily equivalent to the measure of the dynamical degrees of freedom, in the absence of the bath. 

While it is possible to address these questions by numerical simulations, and reconstruct the density that way, what has, really, changed in the last years is that experiments of great precision, that probe both issues, have become possible, particularly in magnetic systems\cite{munzenberg_magnetization_2010}. It is in such a context that the FDR 
 has become of topical interest~
\cite{mitra_spin_2005,safonov_fluctuation-dissipation_2005,coffey_thermal_2012}.

In such systems, since  the noise affects the magnetic field, that makes the spin precess, it is not additive, but multiplicative. While, already, for additive noise, the issue of the ``backreaction'' of the dynamical degrees of freedom on the bath is quite delicate, for multiplicative noise it becomes even more difficult to evade and must be addressed. 

Further complications arise when the fluctuations are colored, namely posses finite intrinsic correlation time\cite{kupferman_it^o_2004,nishino_realization_2015}. In such a situation, no FDR has been unequivocally obtained, that relates the intensity of the  fluctuations to the damping constant~\cite{baiesi_fluctuations_2009}. 

In this note, we wish to study these issues in the context of magnetic systems placed in random magnetic fields, whose distribution can have an auto--correlation time comparable to the time scale defined by the precession frequency. The aim of this communication is to sketch out a route for establishing a FDR in a quite general setting~\cite{aron_symmetries_2010},  that will be shown to be consistent to previous results for magnetic systems, obtained in the limit of white-noise fluctuations, and can be readily adapted beyond this context, especially  for explicit calculations.   A remaining challenge is to obtain  the stochastic equation, that defines the mapping between noise and the dynamical degrees of freedom, that are identified with the spin components of a nanomagnet, and whose solution does, indeed, describe  a normalizable density for the spin configurations. 
 
\section{Gaussian approximation}\label{findimcase}
In order to better grasp the issues at stake, we shall start with a finite number of dynamical degrees of freedom, $s_n^A$. The time index $n$ runs from 0 to $N-1$ and will be identified with the evolution time instant, in the continuum limit; the flavor index $A$ runs from 1 to $N_f$ and labels ``internal'' degrees of freedom--it will label the components of the spin. The summation convention on repeated indices is assumed.

We assume that these dynamical degrees of freedom are immersed in a bath. The bath is described by variables $\eta_n^A$ and is defined by the partition function 
\begin{equation}
\label{ZetaAn}
Z=\int\prod_{A=1}^{N_f}\prod_{n=0}^{N-1}\,d\eta_{n}^{A}\,e^{-\frac{1}{2}\eta_n^A {\sf F}_{AB} {\sf D}^{nm}\eta_m^B}
\end{equation} 
The matrix ${\sf F}$ acts on the flavor indices and the matrix ${\sf D}$ on the ``target space'' indices--that describe the instants in time.  The white noise case corresponds to taking ${\sf D}^{nm}=\delta^{nm}/\sigma^2$. The simplest colored noise case corresponds to taking  ${\sf D}^{nm}=\delta^{nm}/\sigma_n^2$, with not all the $\sigma_n$ equal. Furthermore, if it cannot be put in diagonal form at all, then it describes higher derivative effects. 

The average of a functional ${\cal{F}}$ of the variables $\eta_n^A$ is then well defined as
\begin{equation}
\langle{\cal F}\rangle=\frac{1}{Z}\int\prod_{A=1}^{N_f}\prod_{n=0}^{N-1}\,d\eta_{n}^{A}\,{\cal F}[\eta]\,e^{-\frac{1}{2}\eta_n^A {\sf F}_{AB} {\sf D}^{nm}\eta_m^B} 
\end{equation}
From this expression we may deduce the moments of the degrees of freedom of the bath:
\begin{equation}
\label{moments_bath}
\begin{array}{l}
\displaystyle \left\langle\eta_n^A\right\rangle = 0\\
\displaystyle \left\langle\eta_n^A\eta_m^B\right\rangle = \left[{\sf F}^{-1}\right]^{AB}\left[{\sf D}^{-1}\right]_{nm}
\end{array}
\end{equation}
with the others deduced from Wick's theorem. 
What we notice here is that, for non--diagonal matrices, ${\sf F}$ and ${\sf D}$, the degrees of freedom of the bath that have well--defined properties, i.e. the degrees of freedom that are eigenstates of these matrices, are linear combinations of the $\eta_n^A$. So it makes sense to work in that basis. In this context, the white noise limit corresponds to the case in which ${\sf D}$ is the identity matrix--all components have the same relaxation time. The colored noise case, then can be identified as  that, where ${\sf D}$ is not the identity matrix. 

When we immerse a physical system in such a bath it can happen that the eigenbases of the system and of the bath do not match. 

The map between the degrees of freedom of the bath and the dynamical degrees of freedom is provided by a stochastic equation. For instance, one consider the Landau-Lifshitz-Gilbert equation ${\dot{\bm s}}={\bm\omega}\times{\bm s}+\alpha{\bm s}\times{\dot{\bm s}}+{\overline{\overline{E}}}(s){\bm\eta}$, where the vielbein ${\overline{\overline{E}}}$ contains both an antisymmetric part $\times{\bm s}$ and at least an additional non-zero diagonal element. Because this vielbein is invertible, we can express ${\bm\eta}$ as a function of ${\bm s}$.

To illustrate the procedure, we  start with the case of linear equations: 
\begin{equation}
\label{langevin_map}
\eta_n^A = {\sf f}^{A}_{B}{\sf C}_n^m s_m^B
\end{equation} 
Assuming that the matrices are invertible, we obtain the change of variables (we shall study presently what happens when the matrices have zero modes)
\begin{equation}
\label{Change_of_Var}
 s_n^A = \left[{\sf f}^{-1}\right]^{A}_{B}\left[{\sf C}^{-1}\right]_n^m\eta_m^B
\end{equation}
The Jacobian is a constant that can be absorbed in the normalization of the partition function\cite{zinn-justin_quantum_2011}, so we obtain the partition function for the dynamical degrees of freedom,
\begin{equation}
\label{Zphi}
Z = \int\prod_{A=1}^{N_f}\prod_{n=0}^{N-1}\,d s_{n}^{A}
e^{-\frac{1}{2} s_{n'}^{A'}{\sf f}^B_{B'}{\sf F}_{AB}{\sf f}^A_{A'} {\sf C}_n^{n'}{\sf D}^{nm}{\sf C}_m^{m'}   s_{m'}^{B'}}
\end{equation}
that defines the correlation functions--for the finite-dimensional case the moments--of the dynamical degrees of freedom. The 1--point function vanishes, $\langle s_n^A\rangle = 0$, while the 2--point function is given by the expression 
\begin{equation}
\label{2pointPhi}
\left\langle s_n^A s_m^B\right\rangle = \left[[{\sf f}^{-1}{\sf F}{\sf f}]^{-1}\right]^{AB}\left[[{\sf C}^{-1}{\sf D}{\sf C}]^{-1}\right]_{nm} 
\end{equation}
This is the FDR for the present case, that relates the parameters, ${\sf f}_B^A$ and ${\sf C}_n^m$, of the spin dynamics, with the parameters, ${\sf F}_{AB}$ and ${\sf D}_{nm}$, of the bath. 

\section{When zero modes are relevant}\label{analysis}

Let us now consider the case when the matrices ${\sf f}^A_B$ and/or ${\sf C}_n^m$ have zero modes, a case that is relevant for the physical system studied in this paper. 

The zero modes imply, quite simply, that we cannot replace all of the $\eta_n^A$ by the $s_n^A$, since we cannot invert eq.~(\ref{langevin_map}); we can, only, replace the non--zero modes. The matrices ${\sf f}$ and/or ${\sf C}$ are not of full rank--but they surely have positive rank, otherwise the stochastic map does not make sense. When we replace the non--zero modes, we shall generate quadratic terms in the $ s_n^A$--but, since we do not replace all of the $\eta_n^A$, on the one hand there will be mixed terms, while there will remain the terms quadratic in the $\eta_n^A$, that correspond to the zero modes. When we integrate over the zero modes, the $\eta_n^A$ that we could not express directly as linear combinations of the $ s_n^A$, we shall encounter Gaussian integrals over them that contain terms linear in the zero modes and the $ s_n^A$ already replaced. The result of these Gaussian integrations will be quadratic contributions to the already present $ s_n^A$, that enter in the action with the opposite sign to their coefficients. The system will be stable, if these contributions do not completely cancel the existing ones and will lead to a modification of the FDR. 

Let us see this in action. We shall take $N_f=3$ and  ${\sf f}^A_B=\varepsilon^{A}_{\hspace{2mm}BC}\omega^C$, with $\bm{\omega}$ a fixed vector in flavor space. In the magnetic case it will correspond to the fixed part of the precession frequency. We immediately remark that ${\sf f}^{A}_B$  has one zeromode, along the vector $\bm{\omega}$.  Since this vector is fixed, without loss of generality, we may take it to lie along the $z-$axis: $\bm{\omega}=(0,0,\omega^3)$. 

The stochastic equation, eq.~(\ref{langevin_map}), takes the form
\begin{equation}
\label{stochastic_map_zm1}
\begin{array}{l}
\displaystyle
\eta_n^1 = \omega^3 {\sf C}_n^m s_m^2\\
\displaystyle 
\eta_n^2 = -\omega^3 {\sf C}_n^m s_m^1 
\end{array}
\end{equation}
We may replace these in the partition function for the noise; but we must integrate over $\eta_n^3$ separately. We remark that they do not involve $ s_n^3$, the component of the dynamical degrees of freedom, parallel to the precession vector. 

If ${\sf F}_{AB}=\delta_{AB}$, i.e. the spherical symmetry is imposed, we immediately deduce that the integration over $\eta_n^3$ decouples from the rest and just gives a contribution to the normalization. The partition function for the dynamical degrees of freedom, $ s_n^1$ and $ s_n^2$, is given by the expression
\begin{equation}
\label{Zphysical}
Z = \int\prod_{A'=1}^{2}\prod_{n=0}^{N-1}\,d s_{n}^{A}\,e^{-\frac{\left(\omega^3\right)^2}{2}  s_{n'}^{A'}\left[{\sf C}{\sf D}{\sf C}\right]^{n'm'} s_{m'}^{A'}}
\end{equation} 
There's a subtle point here: the motion of the $A'=1,2$ flavor components is a rotation, with precession frequency $\omega^3$, about the $z-$axis, so the combination, $( s_m^1)^2 + ( s_m^2)^2$ should appear--and it does. Therefore we deduce the FDR for this case, that corresponds to Larmor precession:
\begin{equation}
\label{FDR_Larmor}
\left\langle s_n^A s_m^B\right\rangle = \left(\omega^3\right)^{-2}\left[{\sf C}^{-1}{\sf D}{\sf C}\right]^{-1}_{nm} 
\end{equation}
If the spherical symmetry is not imposed, in flavor space, e.g. ${\sf F}_{AB}=\kappa\delta_{AB} + \lambda_{AB}(1-\delta_{AB})$, we would have had terms linear in $\eta_n^3$, along with the quadratic terms and additional contributions when we would have integrated over the $\eta_n^3$. 

\section{Beyond the Gaussian approximation}\label{NonGauss}
Now let us address the issue of non--linear stochastic maps, also relevant for the Landau--Lifshitz--Gilbert equation. Let us replace eq.~(\ref{langevin_map}) by 
\begin{equation}
\label{nlin_langevin_map}
\eta_n^A = {\sf f}_{(1)B}^{A}{\sf C}_n^{(1)m} s_m^B + {\sf f}_{(2)BC}^A{\sf C}_{n}^{(2)ml} s_m^B s_l^C 
\end{equation}
In this case the Jacobian of the transformation, between the degrees of freedom of the bath and the degrees of freedom that describe the ``interesting'' dynamics, is not a constant:
\begin{eqnarray}
\label{Jacobian_gribov}
J_{nk}^{AB}(s)&\equiv&\frac{\delta\eta_n^A}{\delta s_k^B}={\sf f}_{(1)B}^{A}{\sf C}_{n}^{(1)k}+\nonumber\\
\!\!\!\!&+&\!\!\left[{\sf f}_{(2)BC}^{A}{\sf C}_{n}^{(2)kl}+{\sf f}_{(2)CB}^{A}{\sf C}_{n}^{(2)lk}\right] \!\! s_l^C
\end{eqnarray}
This means that, if it is possible to neglect the zero modes and the concomitant fluctuations in the sign of the determinant, which is true in perturbation theory, the partition function for the spin degrees of freedom is given by the expression
\begin{equation}
\label{Zspin_nlin}
Z =\int\left[d s_n^A\right]\mathrm{det}J_{nk}^{AB}({\bm s})e^{-\frac{1}{2}\eta_n^A({\bm s}){\sf F}_{AB}{\sf D}^{nm}\eta_m^B({\bm s})}
\end{equation}
where the $\eta({\bm s})$ are defined by eq.~(\ref{nlin_langevin_map}). The expression in the exponent contains terms that are  quadratic and quartic in the spin variables. The fluctuation--dissipation relation can then be deduced from the Schwinger--Dyson equations\cite{zinn-justin_quantum_2011},
\begin{widetext}
\begin{equation}
\label{SchwDysEqs}
Z^{-1}\int\,\left[d s_n^A\right]\,\frac{\partial}{\partial s_k^L}\left\{s_{n_1}^{A_1}\cdots s_{n_I}^{A_I}\,\mathrm{det}\,J_{nk}^{AB}({\bm s})\,e^{-\frac{1}{2}\eta_n^A({\bm s}){\sf F}_{AB}{\sf D}^{nm}\eta_m^B({\bm s})}\right\}=0
\end{equation}
\end{widetext}
These relations can be used to generalize eqs.~(\ref{FDR_Larmor}) and express the fact that the spin degrees of freedom are in equilibrium with the bath. The determinant can be introduced into the exponent using anti-commuting variables, that describe the dynamics of the bath\cite{zinn-justin_quantum_2011}. 

It should be stressed that, since the $\eta({\bm s})$ are polynomials in the spin degrees of freedom, once the determinant has been expressed in terms of anti-commuting fields, there is a finite number of parameters that define the dynamics and, thus, enter in the fluctuation--dissipation relation. Indeed, if ${\sf D}_{nm}$ is not the identity matrix, which means that the dynamics is not ultra--local in time, tunneling between configurations implies that the effects of the determinant and it sign will be, inevitably and, thus, implicitly, be generated by the dynamics, therefore it suffices to sample the correlation functions by the action of the spin degrees of freedom. The subtleties of the dynamics are encoded in the relation between the noise fields and the spins, so it is at that point that the zero modes need to be taken into account. There are not any issues of principle, involved, however, precisely because the system is consistently closed~\cite{nicolis2016}.
How to sample the correlation functions will be reported in detail in future work.

\bibliographystyle{apsrev4-1}

\end{document}